# Correlative-causative structures and the 'pericause': an analysis of causation and a model based on cellular biology

Sepehr Ehsani

Department of Laboratory Medicine and Pathobiology, and Tanz Centre for Research in Neurodegenerative Diseases, University of Toronto, Toronto, Ontario M5S 3H2, Canada

*Present address:* Whitehead Institute for Biomedical Research, and MIT Computer Science and Artificial Intelligence Laboratory, Cambridge, Massachusetts 02142, United States

ehsani@csail.mit.edu

October 1, 2013

*The advent of molecular biology has led to the identification of definitive causative factors for a number of diseases, most of which are monogenic. Causes for most common diseases across the population, however, seem elusive and cannot be pinpointed to a limited number of genes or genetic pathways. This realization has led to the idea of personalized medicine and treating each case individually. Nevertheless, since each common disease appears to have the same endpoint and phenotypic features in all diagnosed individuals, the search for a unifying cause will still continue. Given that multivariate scientific data is of a correlative nature and causation is always inferred, a simple formalization of the general structure of cause and correlation is presented herein. Furthermore, the context in which a causal structure could take shape, termed the 'pericause', is proposed as a tractable and uninvestigated concept which could theoretically play a crucial role in determining the effects of a cause.*

**MAIN TEXT**
The main aim in molecular and cellular biology, and in the sciences in general [1,2], is the true *understanding* of cellular and subcellular processes. Furthermore, it is hoped that this understanding would lead to the identification of causal factors of disease and, therefore, pave the way for as-yet-unattained curative treatments (see for example [3]). Although monogenic causes for a number of rare diseases have been identified, the majority of almost all common diseases seem too complex to be caused by one or more specific factors on a population level. This has led to strategies where each patient is considered individually and a case presenting with a common disease is treated as a unique and rare disease. Nevertheless, because an apparently unifying thread seems to connect all cases of a given type of common disease, the need to discover an underlying causative factor leading to the occurrence of, for example, solid tumors or neurodegenerative diseases, persists.

Scientific observations or measurements, however, are all of a correlative (preferably reproducible) nature, subsets of which may be more permissive to characterization as causal associations. Although various methods have been developed to aid in the inference of causal associations from correlative data (for example [4]), such methods rely on *a priori* assumptions based on the experimenter's state of understanding about the problem at hand. To aid in better formulating questions in the domain of causal inference from correlative data, it may be useful to expand on the underlying structures.

**Structures of correlations**
To define certain parameters within correlative relations, a categorization of different correlative possibilities is presented in **Figure 1**. These possibilities have been designated as 'cause-effect (CE)'-null, CE-inherent, CE-complete and CE-incomplete.

The **CE-null** category consists of instances where although it is thought that a change in *A* is leading to a change in *B*, in fact, a hidden variable *x* is being affected, which leads to a concomitant change in *A* and *B*. No cause-and-effect relationship can therefore be inferred from the CE-null category.

If *A always* leads to *B*, *A* and *B* therefore (i) cannot exist independently of each other, (ii) are not two different existences and (iii) cannot form a causative structure. Hence, their singular identity and true representation should be *AB*. This category can be termed **CE-inherent**. An example could pertain to the physical falling of an object, where it cannot be said that the force of gravity *causes* the falling of an object, for the act of falling is an irrevocable consequence of what we define as gravity, and therefore becomes meaningless if thought of independently of gravity. CE-inherent cases are two manifestations of the same entity, and can have no examinable causal relationship.

If the experimenter can be assured that an intended change in *A* is in fact only leading to a change in *A*, and that a subsequent change can be observed in a single facet of *B*, a ***CE-complete*** scenario can be defined and a causal link can be investigated with greater certainty (it should be noted that one or more intermediaries may or may not exist between *A* and *B*). If, however, the effect on *B* is multifaceted, a ***CE-incomplete*** category can be imagined. For example, if increased levels of a protein lead to cell death, but at the same time allow for greater growth of a resistant colony in the same pool of cells, the effect of the protein on cellular toxicity should be further refined so that a particular facet of cellular toxicity (and not toxicity in general) is considered as having a causal relation with the said protein.

**Structures of causes**
Since CE-complete cases are the most amenable to inferring causal relationships, it would be beneficial to examine how a theoretically ideal causative structure would develop over time (**Figure 2**). Starting at $t_1$, the cause *A* is not present in a domain which could influence *B*. At $t_2$ and $t_3$, *A* is present in a domain to influence *B*, but the effect is not accomplished due to the circumstances/context at those time points. At $t_4$, however, the effect can be accomplished and at $t_4+x$, the effect is completed. The course of events is thus changed. Based on this model, a few corollaries can be proposed:

(i) It is rational to assume that cause and effect must follow each other in time [5].
(ii) Correlation is temporally continuous whereas causation is discrete, since there could be significant temporal disconnects in an otherwise continuous causative relation.
(iii) Given that the mere presence of a variable in a causal sphere of a second variable is not sufficient for an effect to be exerted, past or present precedents cannot be solely relied upon to predict the effect of a present cause in the future without more information.
(iv) If a certain experimental paradigm does not capture a CE-complete correlation between two variables, that does not necessarily preclude those variables from demonstrating CE-completeness under a different paradigm and at a different time (for further discussion, see [6,7]).
(v) Discoveries of cause-and-effect in the present are bound to find instances in the future where the causative structures temporarily or permanently break down.

**The 'pericause'**
Based on the model presented above, it is evident that knowledge of the nature of a cause and the nature of the affected variable will be insufficient in truly understanding the correlative-causative structure at hand. To demonstrate this insufficiency, a simple genetic-regulatory-protein network is presented in **Figure 3**. In this example, Protein 1 and Protein 2, encoded and regulated by their respective gene and RNA architectures, lead to Subphenotype 1 and Subphenotype 2, respectively. The subphenotypes together form Phenotype A. Upon a perturbation, in disease, of a component of Protein 2's architecture, Subphenotype 2 is morphed into Subphenotype 2'. Using a numerous-layered fallback system, the cell exerts a computationally efficient disturbance in another protein's architecture, for example Protein 3, to revert Subphenotype 2' to Subphenotype 2. The cell can theoretically rely on innumerable combinatorial strategies to create the most efficient self-imposed corrective disturbance, which may differ from cell to cell even in the same tissue in one organism. This is evidenced by observations that (i) the contribution of many associated genes to a complex phenotype is minor, (ii) knockouts of many genes may not lead to a striking phenotype, and (iii) any given gene may participate in novel and unknown functions (see for example [8,9]).

As demonstrated by the above example, even in the case of a simple genetic network, scenarios could be readily imagined where the effect of one cause on two cells could be too variable and non-reproducible based on the state of the cells' fallback machinery. Even in cases where a cause does have a reproducible effect on a large sample of a given population, the identification of the ultimate cause (the cause of the cause) [10], and the directionality of the cause [11] pose further challenges in understanding the causal structure.

These challenges point to the **context of the cause**, herein referred to as the 'pericause', as a leading determinant in the outcome of a causative relation (**Figure 4**). The pericause can, for example, be imagined as a box containing *B* submerged in water (*A*), preventing the direct contact of *A* and *B* and therefore leading *B* to remain dry. A cause cannot exist independent of the pericause and, in essence, **the pericause determines the probability of *A* being the cause of *B*** across a population of experimental subjects. Although the pericause can theoretically have many components, time is always the major constituent. Although time can never act as a cause, it is ubiquitous in all causal structures based on its shaping of the pericause. In comparing causal structures, however, 'universal' time cannot be assumed to be acting equally on all structures. For example, the scale and potential variability of local time on a cellular level cannot be equated or compared to the local time of human interactions.

**Conclusions**
The pericause may represent an examinable component of correlative-causative structures with the aim of better understanding questions pertaining to natural phenomena which have eluded thorough explanation. Although it is not clear what the full range of examinable pericauses could be, one example of the many candidates in cell biology could be the picosecond vibrations of the plasma membrane [12], possibly acting as a pacemaker in the cell and shaping the effect of various extracellular or genetic/regulatory causes. Moreover, following from this example, it is also not clear which of the current developing technologies in the field of molecular biology focusing on one or more of the domains of the central dogma [13] would be applicable in studying cellular pericauses. Methods to analyze the pericause certainly require careful consideration.

Overall, an immediate goal in this area could be to develop a system of representation to define the parameters of candidate pericauses (and the interactions of those candidate pericauses) in a given problem and to develop hypotheses to probe their validity. It remains to be seen whether such representations could be developed mathematically or if another representative language that can capture different forms of complexity should be developed [14].

**FIGURE LEGENDS**
**Figure 1. Correlative structures.** Correlative data captured by a given experiment can be broken down into categories based on the perceived relationship between the cause (*A*) and the effect (*B*). 'CE' stands for 'cause-effect'.

**Figure 2. Causative structures.** In a theoretical model, the effect of *A* on *B* is not dependent on precedents. *A* causing *B→B'* at $t_4$ cannot necessarily predict the effect of *A* on *B* at a future time, even if *A* remains within *B*'s causative sphere.

**Figure 3. Elusive cause in a genetic network.** Due to the innumerable fallback mechanisms in genetic networks, pinpointing a unifying cause of a disturbance even among a handful of cells is a difficult task.

**Figure 4. The pericause.** The context of the cause (i.e., the 'pericause') can be imagined as shaping all correlative-causative interactions, and thus as a potential avenue for intervention in pathobiology.

**Figure 1**

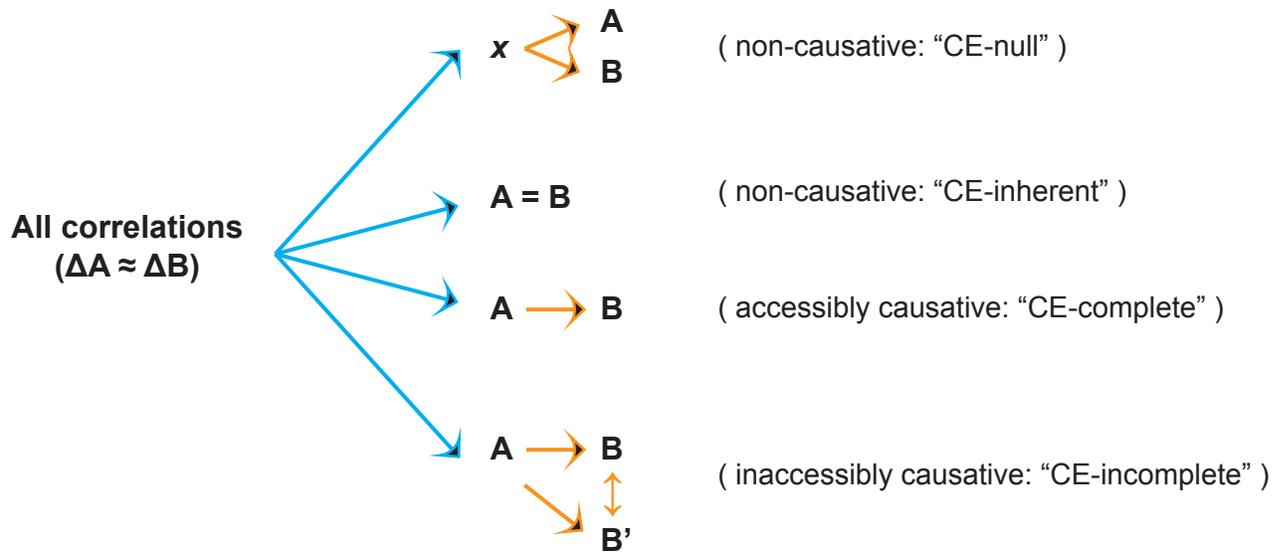

**Figure 2**

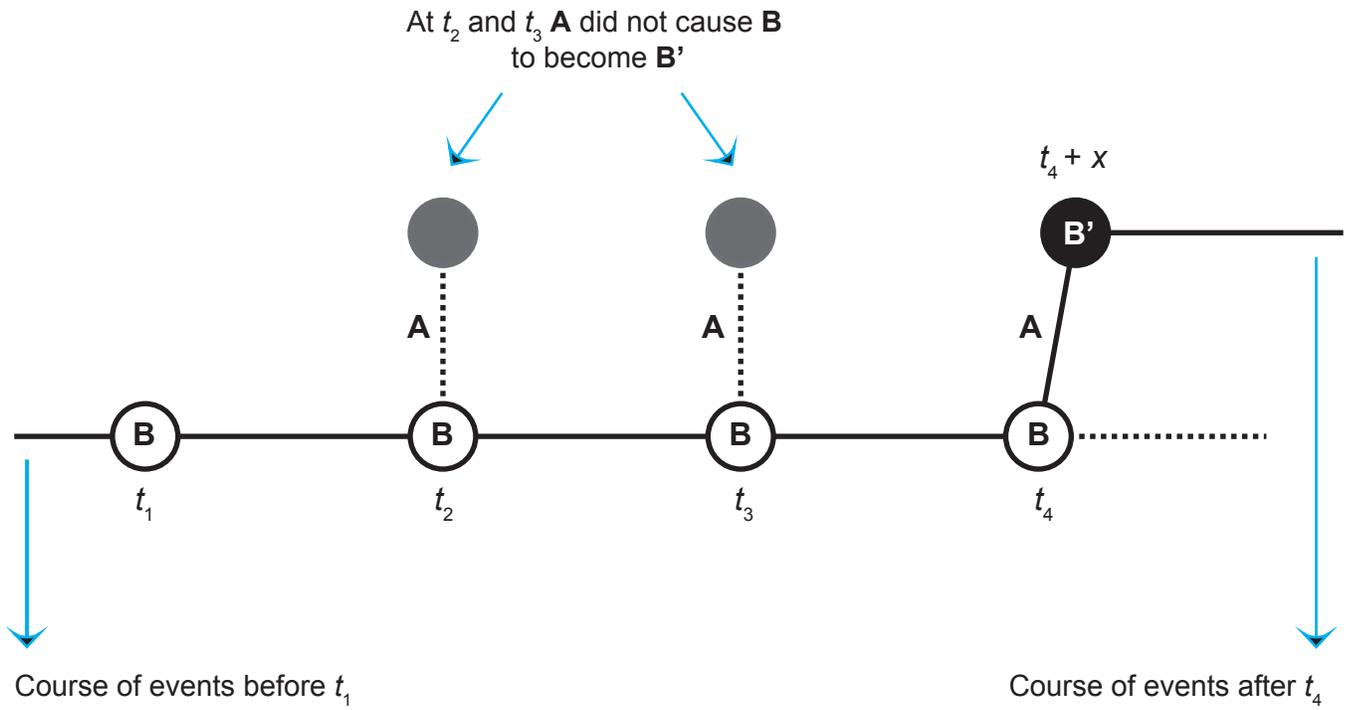

**Figure 3**

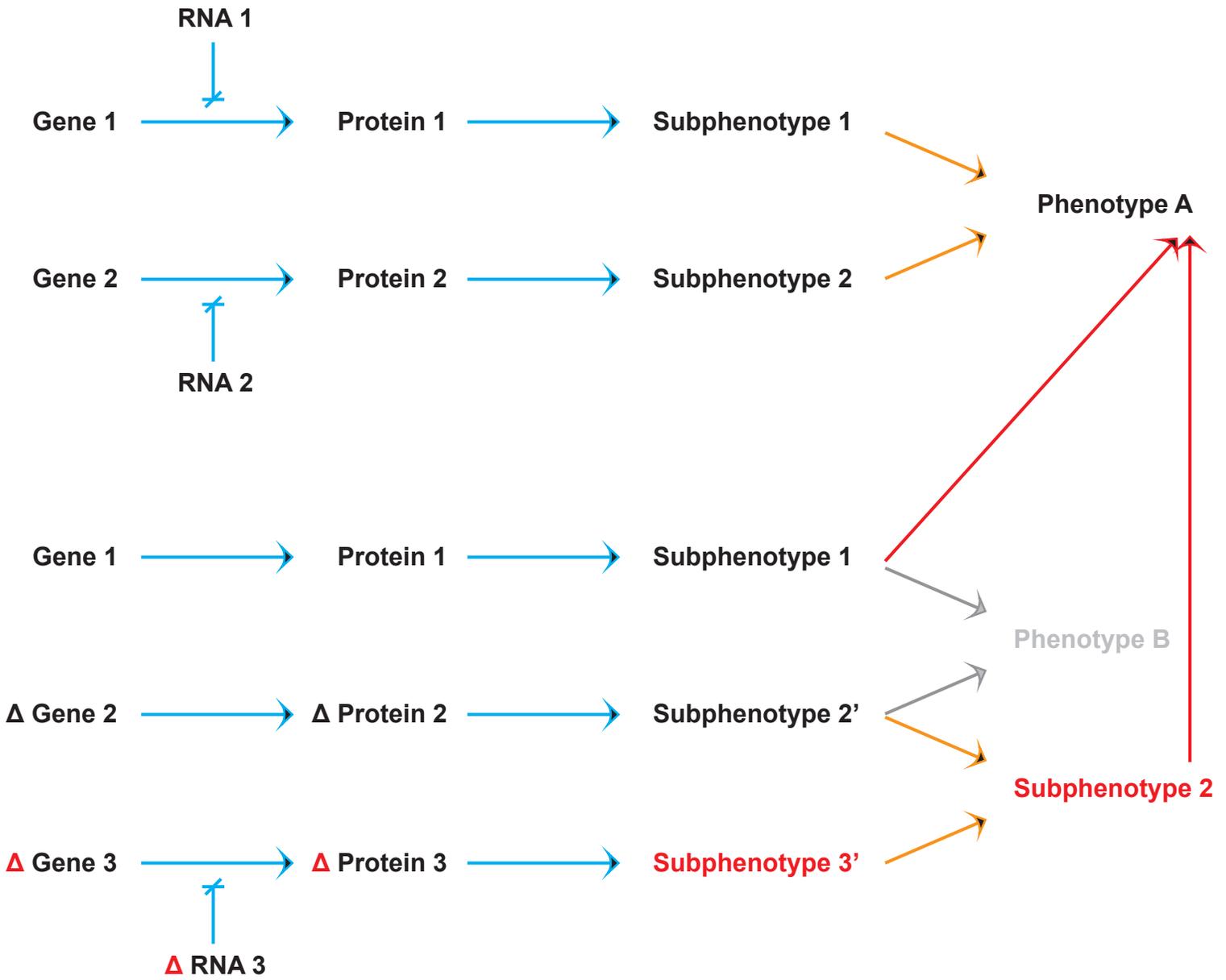

**Figure 4**

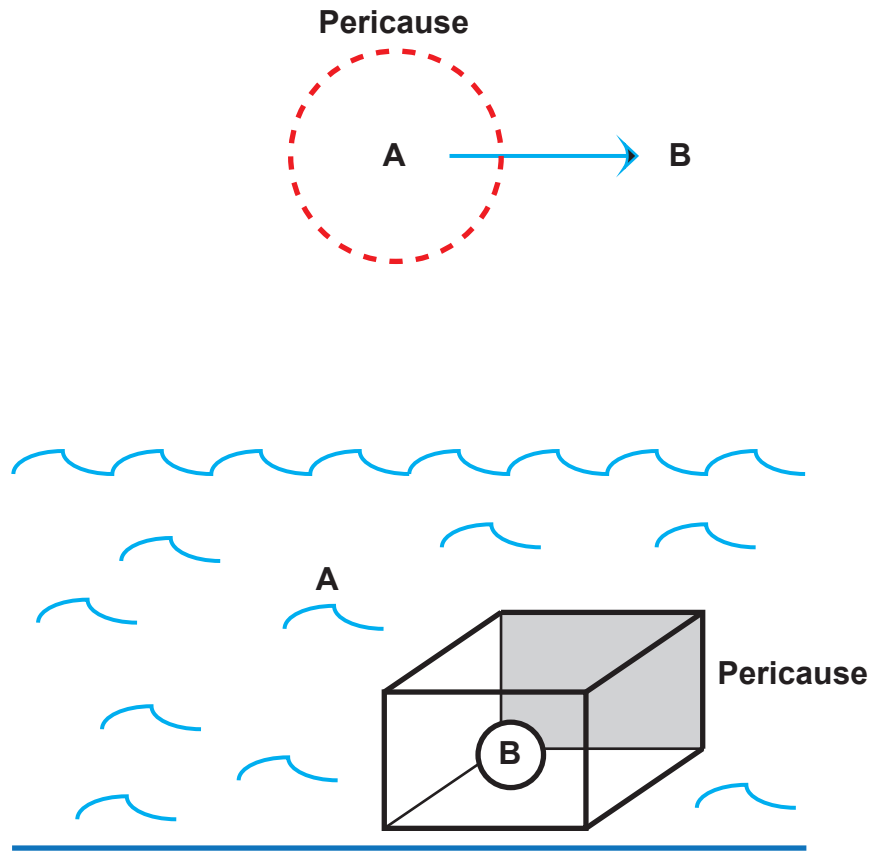